\documentstyle[
preprint
 ,twoside
 ,adbis96
 ,epsf
 ,times]{acmconf}
\newenvironment{namelist}[1]{%
\begin{list}{}
    {
      
      \settowidth{\labelwidth}{#1}
      \setlength{\leftmargin}{1.1\labelwidth}
    }
  }{%
\end{list}}

\author{V.E.Wolfengagen\thanks{also:
Kashirskoe Avenue, 31, Cybernetics Department, Moscow Engineering
Physical Institute, Moscow, 115409, Russia }
              \vspace{1.52mm} \\
{Vorotnikovsky per., 7, bld. 4} \\
{Dept. of Advanced Computer Studies and Information
Technologies}\\
{Institute for Contemporary Education ``JurInfoR-MSU''} \\
{Moscow, 103006, Russia} \\
{\tt vew@jmsuice.msk.ru} }
\addtocounter{footnote}{1}
\title{\bf Objects and their computational framework%
\thanks{This research is granted in part by
Russian  Foundation for Basic Research (project 96-01-01923). The
research of computational models is supported in part by Institute
for
Contemporary Education ``JurInfoR-MSU ''.}%
}
\newtheorem{exm}{\sf Example}[section]

\newtheorem{mtm}{\sf Metateorem}[section]

\begin{document}

\bibliographystyle{alpha}

\maketitle


\begin{abstract}
Most of the {\em object} notions are embedded into a {\em logical}
domain, especially when dealing with a database theory. Thus,
their properties within a {\em computational} domain are not yet
studied properly. The main topic of this paper is to analyze
different concepts of the distinct computational {\em primitive
frames} to extract the useful object properties and their possible
advantages. Some important metaoperators are used to unify the
approaches and to establish their possible correspondences.
\end{abstract}



\section{Introduction}

\subsection{Topics}

This note is inspired with a tremendous variety of the {\em known}
and {\em unknown} to the database theoreticians
notions of an {\em object}~\cite{Wand:89}, \cite{Wan:94}, \cite{Coa:91}.
A traditional approach to understand an object as a database theory
phenomena tends to bring the {\em logical} aspects to
study all its practically necessary and
interesting features~\cite{Beer:90}.

The implementation efforts often violate the object harmony within
any prescribed logical domain especially when discovering some
additional computation effects~\cite{Mano:90}, \cite{Fong:91}.
Under these conditions the idea that only the logic gives a {\em
sound} ground to establish and investigate the objects becomes not
so attractive.

From the intuitive reasons an object is understood as a relatively
self contained and  stable  entity which attaches
both the building blocks
and toolkit
facilities with a possible aim to develop and implement
some kind of the information system~\cite{Rou:76}, \cite{HeCo:89}.

The discussions of the minimal mathematical amount
to encircle the tasks implied are known
and share the same domain:
how to start with the minimal assumptions and restrictions
for generating the most of the known and already used
object features~\cite{Coy:92}.
Thus, the notion of object makes the boundaries
of the current theoretical issues concerning
not only the {\em logical} part of a database theory
but also its {\em computational} part.

\subsection{Related Research}

The research efforts are concentrated on the
main concepts, vital theories-and-foundations, metatheoretical
considerations
to resolve from an object positions
an essence of the database universe
of discourse~\cite{Brod:95}, \cite{Wied:92}.

The advance in the logical studies of an object
is based on the attempts to discover the suitable
mathematical representation, and the {\em individuals}
covered the gap between the intuitive ideas
and the rigorous ground.
The most prominent results have been established when
all the individuals were divided into
{\em actual}, {\em possible} and {\em virtual} ones~\cite{Til:93}.
The notion of possible individual~\cite{Sco:70}
brought forward the {\em schematic} nature of an object
and added more flexibility into the pure logical models.
Thus, a {\em state} of the object was clear represented
and studied in more details giving rise to
data dynamics~\cite{Wo:93}.

The intuitively observed objects were enclosed into some
mathematical universe of discourse, and, e.g., a category
theory~\cite{NeRo:95} is one of the promising candidates to
establish the desired theoretical framework. In a category object
is evaluated with the {\em assignment} which captures the
properties of the computational environment. The changeable
assignments simulate the dynamic effects to manipulate the object
states. The possible invariants add the stability for the objects,
and the triples $<$state, individual, concept$>$ give the sound
basis to use and represent this kind of objects as the basic
building blocks for the target information system~\cite{Wo:93}.

This approach possessed of both the logical and computational
properties. From time to time the more attention was payed to
one of the features than to another.
After a period of the research activity the new kind of
object was discovered - the {\em variable} domain~\cite{Sco:80}.
A hope to deal with the mathematically sound objects
was reached as the {\em functor-as-object}~\cite{Wo:96}.

In turn, the functors were used for ranges of variables
from the logical formulae to simulate the {\em polymorphic types}.
At once, the initially computational idea became more
logical violating the proportion between the counterparts.

The situation implies the backtracking to revise the necessary
meaning which is put into the initial notion of an object.
The attention is turned back to develop a conceptual framework
that is adequate for characterizing the computational features
of the {\em object}~\cite{Yah:87}.
The basic task is to assemble existing computational cases
and concepts to form a suitable framework. The descriptions
thus obtained provide a perspective from which to view
possible advantages and disadvantages for computations
with the objects as they are~\cite{Gil:90}.

\subsection{Aims}

The descriptions of various kinds of computation constitute part
of a framework which can accommodate {\em combinatory} logic. In
an initial (and pure) theory the {\em combinators-as-objects} have
been used to show that bound variables are unnecessary in systems
of logic. When {\em formulae} are used to restrict an object
properties then bound variables mark the places in a formula that
are affected by some {\em metaoperator}, in particular, a
quantifier.

In their mathematical origin, combinators are objects
which express rules for manipulating other objects
(and combinators among them).
The generality of an approach to represent `applied' objects
by `pure' objects can be conditionally restricted
to capture the needed features and computational effects.

Some of the following questions are of interest
to many researchers and to the author also:

\begin{namelist}{{\tt (4)} {\it {   }}}
\item {(1)} Non-formal ideas concerning `object'.
\item {(2)} Do the known formal theories of objects really
fruitful to capture an `object' intuitive reasons amount?
\item {(3)} Data model based on computations with the objects:
        is it a phantom or the desirable means?
\item {(4)} Base of data vs. database: the basis property
for computations with the objects.
\item {(5)} Inductive classes: generating a variety of (possible) objects
        (which are schematic).
\item {(6)} Object computation style: what are the images?
\item {(7)} Imposing an algebraic properties and structures.
\item {(8)} Mechanisms: inheritance, encapsulation, polymorphic types, etc.
\item {(9)} Conceptual shell and encoding the realistic objects by the
              $<$concept, individual, state$>$ triples.
\item {(10)}  Relativity of object dynamics.
\end{namelist}

The present paper not obviously covers all the troubles.
The mathematical trends and citations for the
contemporary research activity in combinatory logic,
$\lambda$-calculus, and category theory are omitted.
Only a few of the related topics are used reflecting
the current interests of the author.

The paper is divided into four sections.
Section 1 gives some needed encircling of the object topics.
Section 2 contains the pre-ranging the expressions with
the objects via different general metaoperators
setting up the initial conceptual framework.
Section 3 illustrates the particular computational frameworks
within the initial one.

The first two sections are independent of combinatory logic
(and purification of combinator-as-object). The third section
deals with the particular
{\em applicative computational systems}, i.e. an application
metaoperator is significant. The functional abstraction
as a metaoperator is unnecessary to use -
its effects may be assembled into combinators.

\section{Computational Consideration}

\subsection{Non-Formal Ideas Concerning `Object'}

The natural way to represent the ideas involved is to verify
if are there any atomic, and simplest, entities.
Those entities are to be used to generate the derived entities that are
built from other, and less complicated entities.
More suitable way is to propose different modes to establish and use
entities.
First of them is as follows:
\begin{namelist}{{\tt (4)} {\it {   }}}
\item {(1)}~the researcher starts with the simplest
entities and expands them to generate more complicated ones;
\item {(2)}~the relation of {\em expansion} should be
established to make a linkage
between the initial and target objects.
\end{namelist}

To the contrary the second approach gives another way:
\begin{namelist}{{\tt (4)} {\it {   }}}
\item {(1)}~the researches takes an entity as it is and makes an attempt
to reduce it to less complicated entities;
\item {(2)}~the relation of {\em reduction} is used to make a linkage between
the initial and target objects.
\end{namelist}

In case when both expansion and reduction are used simultaneously
the relation of conversion is said to be used.

\subsection{Preliminary Remarks}

An object in mathematics, as a rule, needs the purely abstract
notion to avoid possible ambiguities. The fruitfulness of this notion
depends on the pragmatic sense of the corollaries being extracted.
The distance between the notion of objects in mathematics and
in computer science is even more than the gap between
pure and applied theory. E.g., in applications some kind of logic
may be presupposed and used to fix the useful properties of the
intuitively observed objects. To the contrast, pure and rigorous
consideration does not deal with any presupposed logic to avoid
excessive restrictions. Instead this, the metatheoretical framework
is by default selected to fix the properties of the mathematical tool
under development. Hence, the first remark is as follows:
an initial metatheoretical framework is some kind of pre-logic, at least,
with the (potential) computational properties.
First of all, this means the possibility does exist to built
the usual constructs, e.g., variables, constants, sets,
functions and functional spaces etc. Note that the needed truth values
are to be generated as the specific objects.

On the other hand the known essential computational property is heavy
based on the notion of substitution. Indeed, an everyday
computer science practice involve various replacement strategies
of some parameters by the other parameters or values.
Thus, an importance of substituting process is clear understood,
deeply studied and not yet completed even in the research area.
The main idea is to promote the restricted substitutions
to generate the applied theories of objects.
Being unrestricted, the substitution process directly leads to
the higher order theories -- and to interesting and less understood operators.

\subsection{Ranging the Objects: Metaoperators}

The restriction arises very naturally in different approaches.
A typical way to construe the weak restriction is to enable
the correspondences between objects, e.g., as follows:

\begin{center}
{\em
 Operator($\cdot\ \cdot$): object $\times$ range $\to$ object
}
\end{center}

Here: the {\em operator} acts on an {\em object} which is restricted
by the {\em range}.
This kind of operators often in referred to the intention operators
or metaoperators.

Note that the origin of the initial consideration of the entities
needs from the very beginning some suitable constructs that
individualize the set of properties by the objects.
For convenience they are referred as the individuals.

The last two decade research activity tends to separate
the class of individuals into subclasses,
so the actual, or existing, potential, or possible, and virtual individuals
are distinctly extracted and studied in part.
Next, the correspondences between the actually existing,
potential and virtual entities must be established.

The sensitivity of this separation-and-correspondence depends on
an expressive power of the metamathematical framework.
This is a branch point when the initial homogeneous metatheory is separated
into syntax and semantics. Thus, the constructions containing
the objects are to be evaluated to compute the values of expressions.
An evaluation mapping results to the metaoperator

\begin{center}
{\em
Val($\cdot\ \cdot$): source-object $\times$
assignment $\to$ target-object
}
\end{center}

Here an {\em assignment} marks the context switch,
where the source-object is to be evaluated to result in its value,
or target-object.

\section{Preserving the Computational Potentiality}

\subsection{Involving an Abstraction Metaoperator}

Start with the initial amount of entities: possibly, infinite set
of variables and constants. All the consideration deals mainly with
the notion of function $f$ which corresponds, at least,
one object $f(x_1, x_2, \dots , x_n)$, its value, to $n$-tuple
of objects $x_1, \dots, x_n$, its arguments, which in turn
may be functions in the current sense. For convenience to refer to
the distinct arguments of $n$-ary function the abstraction metaoperator

\begin{center}
{\em $(\lambda\cdot.\cdot)$: variable $\times$ source-object $\to$
target-object }
\end{center}
is established. \\
Here: $\lambda x_1.f(x_1,\dots,x_n)$ determines
that $f(x_1,\dots,x_n)$ is the function of $x_1$;
in $\lambda x_1.f(x_1,\dots,x_n)$ the prefix `$\lambda x_1$' abstracts
the function $\lambda x_1.f(x_1,\dots,x_n)$
of the expression $f(x_1,\dots,x_n)$. The clear abbreviation
for arbitrary function
$F$ gives
\[
\lambda x_1 \dots x_n.F=(\lambda x_1.
          (\lambda x_2.(\dots (\lambda x_n.F)\dots )))
\]
as an $n$-placed multiabstraction.
To determine objects by the $\lambda$-notation the additional
metaoperator of application is needed, and

\begin{center}
{\em
($\cdot\ \cdot$): object $\times$ object $\to$ object
}
\end{center}
saves the writing efforts.

The intuitive reasons to use abstraction and application notation,
before giving the precise definition of an object are as following.
In general, $\lambda$-expression, or $\lambda$-term is known
as an unary function which values and arguments may in turn
be the functions. Every variable represent an arbitrary unary function,
and $(F G)$ is the result of applying the function $F$ to argument $G$.
Whenever $F$ contains (free) occurrences of $x$,
the $(\lambda x.F)$ represents the function, where its value
for argument $A$ results from the substituting $A$ instead of $x$ into $F$.

Now the class of objects is generated by induction on their complexity,
namely:
\begin{namelist}{{\tt (4)} {\it {   }}}
\item {($i$)}~both variables and constants are the objects;
\item {($ii$}~for objects $F$, $G$ their application $(F G)$ is an object;
\item {($iii$)}~for object $F$ and variable $x$ the abstraction
$(\lambda x.F)$ is an object.
\end{namelist}

The definition above has the `side effect':
set of variables becomes heterogeneous because of binding properties
of the $(\lambda \cdot.\cdot)$-operator.
This effect is clearly observed by an attempt to determine the substitution:
for any objects $F$, $G$ and variable $x$ an effect
$[G/x]F$ of replacing every free occurrence of $x$ in $F$ by $G$ is given
by induction on complexity of $F$:
\begin{namelist}{{\tt (4)} {\it {   }}}
\item {($i$)}~$[G/x]x=G$;
\item {($ii$)}~$[G/x]a=a$ for atomic $a$ and $a \ne x$;
\item {($iii$)}~$[G/x](F_1 F_2)=([G/x]F_1)([G/x]F_2$;
\item {($iv$)}~$[G/x](\lambda x.F)=\lambda x.F$;
\item {($v$)}~if $y \ne x$ and ($y \notin G$ or $x \notin F$)
     then $[G/x](\lambda y.F)=(\lambda y.[G/x]F)$ else
       if $y \ne x$ and ($y \in G$ and $x \in F$)
          then $[G/x](\lambda y.F)=(\lambda z.[G/x][z/y]F)$.
\end{namelist}

Here $z$ is a new variable not included neither in $G$ nor in $F$.
The last step $(v)$ of induction gives the distinction between
free and bound variables.

The primitive frame of the
$(\cdot\ \cdot)$+$(\lambda \cdot.\cdot)$-metaoperators generates
an equational theory of objects below referred as
$(\cdot\ \cdot)$+$(\lambda \cdot.\cdot)$-theory.

{\em Axioms}. \\
$(\alpha)\ \lambda y.F = \lambda v.[v/y]F$ if $y$ is not bound in $F$
and $v$ is both not free and not bound in $F$; \\
$(\beta)\ (\lambda x.F)G = [G/x]F$; \\
$(\rho)\ F = F$. \\

{\em Rules}. \\
$(\mu)\ F = F' \Rightarrow G F = G F'$; \\
$(\nu)\ F = F' \Rightarrow F G = F' G$; \\
$(\xi)\ F = F' \Rightarrow \lambda x.F = \lambda x.F'$; \\
$(\tau)\ F = G$ and $G = H \Rightarrow F = H$; \\
$(\sigma)\ G = H \Rightarrow H = G$; \\
$(\eta)\ \lambda x.Fx = F$ when $x$ is not in $F$.

\begin{exm}
Let $[x,y] = \lambda r.rxy$ be the ordered pair and
$F$~$=(\lambda x.x[4,$ $(\lambda x.x)3])+$. The direct computation within
this equational theory results in
$F$ $= (\lambda~x.x[4, (\lambda x.x)3])+$ $= +[4,(\lambda x.x)3]$
$= +[4,3]$.
Besides that, if `$+$' is understood as the addition in arithmetic
and `$4$' and `$3$' are natural numbers then
$+[4,3] \Rightarrow 4 + 3 \Rightarrow 7$.
\end{exm}

The question arises: is the $(\lambda \cdot.\cdot)$-abstraction
metaoperator necessary needed in a theory of objects?
The answer below is negative.

\subsection{Avoiding an Abstraction Metaoperator}

The formal system without an abstraction metaoperator
does exit. But this avoiding leads to some problem
with encapsulation. Even more, the direct consideration
generates the {\em combinatory code} with a lot of
encapsulated objects.

Start with the same as above initial amount of entities:
possibly, infinite set of variables and constants.
The set of constants contains combinators $\cal I$, $\cal K$, and $\cal S$.
In addition, metaoperator $(\cdot\ \cdot)$ of application is used.
An inductive class of objects is generated as follows:
$(i)$~both variables and constants are the objects;
$(ii)$~for objects $F$, $G$ their application $(F G)$ is an object.
Combinator is an object that contains only $\cal I$, $\cal K$, and $\cal S$.

The primitive frame of the
($\cal I$, $\cal K$, $\cal S$) with the $(\cdot \cdot)$-me\-ta\-ope\-rator,
as above, generates an equational theory referred as
($\cal I$, $\cal K$, $\cal S$)$+(\cdot \cdot)$-theory.

{\em Axioms. } \\
For any objects $X$, $Y$, $Z$: \\
({\bf I}) ${\cal I} X = X$; \\
({\bf K}) ${\cal K} X Y = X$; \\
({\bf S}) ${\cal S} X Y Z = X Z(Y Z)$; \\
($\rho)\ X = X$.

{\em Rules.} \\
For any objects $X$, $X'$, $Y$, $Z$: \\
$(\mu)\ X = X' \Rightarrow Z X = Z X'$; \\
$(\nu)\ X = X' \Rightarrow X Z = X' Z$; \\
$(\tau)\ X = Y$ and $Y = Z \Rightarrow X = Z$; \\
$(\sigma)\ X = Y \Rightarrow Y = X$; \\
$(ext)$ if $X V = Y V$ for any object $V$, then $X = Y$.

\begin{exm}
Let $\oplus$ be the curried version of addition,
i.e. $+[4,3]$ $= [4,3]\oplus$ $= \oplus 4 3$
$= (\oplus 4) 3$ and the last object is equal to $7$
by the rules of arithmetic.
Let ${\cal S}({\cal S} {\cal I}({\cal K} \cdot))$
               $({\cal S}({\cal K}{\cal I})({\cal K} \cdot ))$
be the object where dots indicate the missed parameters, so that
\begin{center}
\begin{math}
{\cal S}({\cal S} {\cal I}({\cal K} \cdot))
               ({\cal S}({\cal K}{\cal I})({\cal K} \cdot ))
 : object \times object \to object.
\end{math}
\end{center}
The missed parameter can be called as the {\em encapsulated} objects
whenever the {\em host} object is observed as a kind of context.
The encapsulation of the numbers result in, e.g., the computation as follows:

\begin{math}
{\cal S}({\cal S} {\cal I}({\cal K} 4))
               ({\cal S}({\cal K}{\cal I})({\cal K} 3)) \oplus =\\
= ({\cal S}{\cal I}({\cal K} 4)\oplus)({\cal S}({\cal K I})
                                ({\cal K} 3)\oplus) \\
= ({\cal I} \oplus)({\cal K} 4 \oplus)({\cal S}
                     ({\cal K I})({\cal K} 3)\oplus) \\
= \oplus ({\cal K} 4 \oplus)({\cal S}({\cal K I})({\cal K} 3)\oplus) \\
= \oplus 4 ({\cal S}({\cal K I})({\cal K} 3)\oplus) \\
= \oplus 4 ({\cal K I} \oplus)({\cal K} 3 \oplus) \\
= \oplus 4 ({\cal I}({\cal K} 3 \oplus)) \\
= \oplus 4 3 \Rightarrow 7.
\end{math}

\end{exm}

\subsection{Equivalence of $(\cdot\ \cdot)$+$(\lambda\cdot.\cdot)$-
      and $({\cal I}, {\cal K}, {\cal S})$+$(\cdot\\cdot)$-Theories}

The known result in a theory of applicative computations is
an equivalence of $(\cdot\ \cdot)$+$(\lambda \cdot.\cdot)$-
and $({\cal I}, {\cal K}, {\cal S})$ $+(\cdot\ \cdot)$-theories.
Thus, both the theories of objects deal with the same task
and similar ideas concerning an object.
It means that the $(\cdot\ \cdot)$+$(\lambda\cdot.\cdot)$-object
({\em source} object) can be represented by the
$({\cal I}, {\cal K}, {\cal S})+(\cdot\ \cdot)$-object ({\em target} object).
Even more, the set \{~${\cal I}$, ${\cal K}$, ${\cal S}$~\} is
the computational basis because of the following Metatheorem.

\begin{mtm}
For arbitrary
$(\cdot\ \cdot)$+$(\lambda\cdot.\cdot)$-objects $P$, $Q$
the following is valid: \\
$(i)~\lambda x.x = {\cal I}$; \\
$(ii)~\lambda x.P = {\cal K} P$ if $x$ is not free in $P$; \\
$(iii)~\lambda x.P Q = {\cal S}(\lambda x.P)(\lambda x.Q)$.
\end{mtm}

\begin{exm}
For $F = (\lambda x.x 4((\lambda x.x)3))\oplus$
the following computation gives the assembling into the basis:

\begin{math}
(\lambda x.x 4((\lambda x.x)3))\oplus = \\
= {\cal S} (\lambda x.x 4)(\lambda x.((\lambda x.x)3))\oplus  \\
= {\cal S} ({\cal S} (\lambda x.x)(\lambda x.4))({\cal S}
       (\lambda x.(\lambda x.x)) (\lambda x.3))\oplus \\
= {\cal S} ({\cal S} {\cal I} ({\cal K}  4))({\cal S}
         (\lambda x.{\cal I} )({\cal K}  3))\oplus \\
= {\cal S} ({\cal S} {\cal I} ({\cal K}  4))({\cal S}
         (\lambda x.{\cal I} )({\cal K}  3))\oplus \\
= {\cal S} ({\cal S} {\cal I} ({\cal K}  4))({\cal S}
           ({\cal K} {\cal I} )({\cal K}  3))\oplus. \\
\end{math}

\end{exm}

\subsection{Type Checking}

The natural way to generate functional spaces by the $\lambda$-abstractions
reflects an idea of {\em type} assignment. The type assignment needs
to modify an existing set of objects which are understood as terms.
Before applying modification the set of types is to be determined.
First of all some basic types are assumed to exist, and each of them
represents some set. For instance, the basic type $N$ represents the set
of natural numbers.
The set of types is defined by induction on their complexity:

\begin{namelist}{{\em (iii)} {\it {}} }
\item[{\em (i)}{}]~every basic type is a type;
\item[{\em (ii)}{}]~if $\alpha$ and $\beta$ are the types then
$(\alpha \to \beta)$ is a type.
\end{namelist}

The following properties are presupposed:
types $(\alpha \to \beta)$ are distinct from the basic types,
and $(\alpha \to \beta) = (\alpha' \to \beta')$
implies $\alpha = \alpha'$ and $\beta = \beta'$.
The type $(\alpha \to \beta)$ has the sense of
``the functions from $\alpha$ to $\beta$'' to represent the set
of functions from the set represented by $\alpha$ to the set
represented by $\beta$. An exact set of the denoted functions
depends on the context where typed combinators or $\lambda$-terms are used.
When it is determined, every type represents the set of individuals
or functions.
For simplicity, the terms are identified by the $\lambda$-abstractions.

As usually, for every type $\alpha$ an infinite set of variables
$v:\alpha$ does exist and $\alpha \ne \beta$ implies
$v : \alpha \ne v : \beta$. In accordance with previous consideration,
let $\lambda x.F$ be a primitive term generating operation.

Thus, the typed $\lambda$-terms are defined as follows:

\begin{namelist}{{\em (iii)} {\it {}} }
\item[{\em (i)}{}] all the variables $v : \alpha$ and constants $c : \delta$
are the typed $\lambda$-terms with the types $\alpha$, $\delta$ respectively;
\item[{\em (ii)}{}] for objects $G : (\alpha \to \beta)$ and
$H : \alpha$ their application $(G H)$ has the type $\beta$;
\item[{\em (iii)}{}] for the variable $v : \alpha$ and the object $G : \beta$
an abstraction $(\lambda v.G)$ is the term of type $(\alpha \to \beta)$.
\end{namelist}

This definition implies that every typed $\lambda$-term has the unique type.
Below $(ii)$ is referred as {\bf F}-rule, or ({\bf F})
and $(iii)$ as $\lambda$-rule, or ($\lambda$).

\begin{exm}[composition, ${\cal B}$]

Let ${\cal B}$ be determined by ${\cal B} = \lambda xyz.x (y z)$.
Suppose that the type of $\lambda xyz.x (y z)$ is $(\alpha_1 \to \beta)$
(according to ($\lambda$-rule)). By ($\lambda$) again an initial object
$\lambda x(\lambda yz.x (y z))$ is separated into left part
$x : \alpha_1$ and the right part
$\lambda y(\lambda z.x (y z)) : \beta = \beta_1 \to \beta_2$.
By ($\lambda$) this object is separated into
$y : \beta_1$ and $\lambda z.x (y z):\beta_2=\gamma_1 \to \gamma_2$.
By ($\lambda$) it is separated into $z : \gamma_1$ and $x (y z) : \gamma_2$.

The next step is based on ({\bf F}-rule), and $x (y z) : \gamma_2$
is separated by ({\bf F}) into $x : \delta_1 \to \gamma_2$ and
$y z : \delta_1$.
In turn by ({\bf F}) the object $y z : \delta_1$ is separated into
$y : \Delta \to \delta_1$ and $z : \Delta$. No compounds are observed.

The set of eight types $\alpha_1$, $\beta$, $\beta_1$, $\beta_2$,
$\gamma_1$, $\gamma_2$, $\delta_1$, $\Delta$ is generated.
The type uniqueness for an object implies the following
type equations:
\begin{math}
\alpha_1=\delta_1 \to \gamma_2, \\
\beta_1=\Delta \to \delta_1, \\
\gamma_1=\Delta, \\
\beta=\beta_1 \to \beta_2, \\
\beta_2=\gamma_1 \to \gamma_2.
\end{math}
Hence, the following chain of equalities is to be inferred:
\begin{math}
\alpha_1 \to \beta = \\
= \alpha_1 \to (\beta_1 \to \beta_2) \\
= \alpha_1 \to ((\gamma_1 \to \delta_1) \to (\gamma_1 \to \gamma_2)) \\
= (\delta_1 \to \gamma_2) \to ((\gamma_1 \to \delta_1) \to
            (\gamma_1 \to \gamma_2)).
\end{math}

At the final stage assuming $\delta_1 = b$, $\gamma_2 = c$,
and $\gamma_1 = a$ the derivation results in
${\cal B} : (b \to c) \to ((a \to b) \to (a \to c))$,
where ${\cal B}$ is the combinator of composition.
\end{exm}

This type generating procedure can be implemented with
more or less difficulties. Note that the type checking of
the applicative forms needs no preliminary transformation
of the initial object.

\subsection{Computations in a Category}

Different ways to construe the computation in a category are
observed. Due to~\cite{CoCuMa:85} a categorical abstract machine
became a tool to compile the initial programm into machine
instructions. An advanced study~\cite{Wo:96} is based on the
object-oriented solution to involve the functor-as-object, and the
flexible data models are to be extracted.

\subsubsection{Combinatory Representation}

Traditionally, the set of combinators is fixed
to represent the machine instructions
by the objects.

Let $\{\varepsilon,\ \Lambda,\ <\cdot,\cdot>,\ [\cdot,\cdot],\ \circ,\ Id\}$
be the set of combinators, where:
\[
\begin{array}{lcl}
\varepsilon & : & (B \to C) \times B \to C;                \\
\Lambda     & : & (A \times B \to C) \to (A \to (B \to C)).
\end{array}
\]
Let $[\cdot,\cdot],$ $<\cdot,\cdot>$ be the abbreviations with the meaning
that $[x,y] = \lambda r.rxy$ ({\em pairing} combinator)
and $<f,g> = \lambda t.[f(t),g(t)] = \lambda t.\lambda r.r(f(t))(g(t))$
({\em coupling} combinator) respectively.
This means that both of them are equipped with the
{\em first projection} $Fst$ and the {\em second projection} $Snd$
where $Fst : A \times B \to A$ and $Snd : A \times B \to B$.
For arbitrary mappings $h : A \times B \to C$ and
$k : A \to (B \to C)$ the following equations are valid:
\[
\begin{array}{lcl}
h & = & \varepsilon \circ <(\Lambda h) \circ Fst, Snd>, \\
k & = & \Lambda(\varepsilon \circ <k \circ Fst, Snd>
\end{array}
\]
and thus determine the {\em cartesian closed category} (c.c.c.).
Computations in c.c.c are known and under development in modern
computer science composing the theory of computations.
The definitions by the objects are straightforward:
\[
\begin{array}{lcl}
\Lambda & = & \lambda h.\lambda xy.h([x,y]), \\
D       & = & [\cdot,\cdot] = \lambda xy.\lambda r.rxy,
\end{array}
\]
etc.

\subsubsection{$\lambda$-Representation}

The $\lambda$-abstractions lead to a direct substitutions
to obtain the meaning of expression.
The elegance of computations becomes even more when the
de Bruijn encoding is used. The de Bruijn code indicates
the depth of binding the variables within $\lambda$-expressions,
i.e. the bound variable is replaced by the number of `$\lambda$'
symbols between this variable and the binding `$\lambda$' excluding
this last `$\lambda$' from the account. For instance,
the object $\lambda y.(\lambda xy.x)y$ is encoded by
$\lambda.(\lambda \lambda.{\underline 1}){\underline 0}$.
\begin{exm}
$F = (\lambda x.x[4,(\lambda x.x)3])+$
$ = (\lambda.{\underline 0}[4,(\lambda.{\underline 0})3])+$.
\end{exm}

\subsubsection{Evaluation and Environment}

The main question is to determine the meaning of expressions,
and this depends on the associated values and identifiers,
i.e. on the environment. The usual set of semantic equations
reflects the idea when applying function to its argument
is represented by the order of writing. Thus, the symbol
of argument follows the symbol of function.

\subsubsection{Semantic Equations}

The semantic equations (cf.~\cite{CoCuMa:85}, \cite{Wo:96})
illustrate an idea of context dependent evaluations. Thus, $\rho$
below is the desired context, and this context sensitivity
controls the flow of computations. In case when applicative
computations are used, the resulting set of equations become
extremely transparent:

\[
\begin{array}{lcl}
\|x\|\rho           & = & \rho (x); \\
\|c\|\rho           & = & c;        \\
\|(M N)\|\rho       & = & (\|M\|\rho)(\|N\|\rho); \\
\|\lambda x.M\|\rho d & = & \|M\|([d/x]\rho)
\end{array}
\]
where $\rho$ is an {\em environment},
$\rho(x)$ is the value of $x$ under the environment $\rho$,
$c$ is a constant denoting the value which is constant also
(according to the usual mathematical practice),
$[d/x]\rho$ is the environment where all the $x$ free occurrences
are replaced by $d$.

In general, computation with de Bruijn notation is analogous
to those with the usual combinators when the set of rules
is slightly modified. Some set of rules and agreements is to accommodate
the usual $\lambda$-expressions to de Bruijn encoding.
Let an environment be represented by
$$ \rho = [\dots [(),w_n]\dots, w_0], $$
where the value $w_i$ is associated to the de Bruijn's code ${\underline i}$.
This is a strict restriction. The environments where
an expression is evaluated are the mathematical structures but not arrays.
This choice is due to the efficiency conditions.
First of all this restriction leads to the simplified computation description:
\[
\begin{array}{lcl}
\|{\underline 0}\|[\rho , d]            & = & d;  \\
\|{\underline {n+1}}\|[\rho , d]        & = & \|{\underline n}\|\rho ; \\
\|c\|\rho                               & = & c; \\
\|(M N)\|\rho                           & = & (\|M\|\rho)(\|N\|\rho);  \\
\|\lambda .M\|\rho d                    & = & \|M\|[\rho , d].
\end{array}
\]
The values are not only of self interest but they are interesting
from the supported computations. From the combinatory view,
e.g., the meaning of $(M N)$ is the combination of $M$ and $N$.
Thus, the following three combinators:
\begin{center}
${\cal \$}$ of arity $2$, $\Lambda$ of arity $1$, and $'$ of arity $1$
\end{center}
along with the infinite set of combinators $n!$ in a sense of:
\[
\begin{array}{lcl}
\|{\underline n}\| & = & n!; \\
\|c\|\rho          & = & c; \\
\|(M N)\|          & = & {\cal \$}[\|M\|,\|N\|]; \\
\|\lambda.M\|      & = & \Lambda (||M||)
\end{array}
\]
are to be established.

The equations above generate the translation of semantic equations
to the purely syntactic ones:
\[
\begin{array}{lcl}
0![x,y]           & = & y; \\
(n+1)![x,y]       & = & n!x; \\
(' x)y            & = & x;  \\
{\cal \$}[x,y]z   & = & xz(yz); \\
\Lambda (x)yz     & = & x[y,z].
\end{array}
\]
These equations are similar to ${\cal SK}$-rules:
the first three of them indicate the property to
suppress an argument (like properties of ${\cal K}$),
the fourth rule is the non-curried version of rule for ${\cal S}$,
the fifth rule is exactly the currying, i.e.
transformation of a function of two arguments into the function
of the first argument which in turn is the function of the second argument.

An additional couple combinator bring more harmony into the
syntactical equations:
$$ \|[M,N]\| = <\|M\|,\|N\|> $$
(this will be shown). This combinator is equipped with the selectors,
or projections $Fst$ and $Snd$. Also consider the composition
`$\circ$' and the additional command $\varepsilon$.
The objects ${\cal \$}[\cdot,\cdot]$ and $n!$ are the abbreviations
for `$\varepsilon \circ <\cdot,\cdot>$' and `$Snd \circ Fst$' respectively,
where $Fst^{n+1} = Fst \circ Fst^n$.
Now all is prepared to write down the syntactical equations.

\subsubsection{Syntactical Equations}

The merging of the previously given sets of rules results in the following:
\[
\begin{array}{llcl}
(ass)      & (x \circ y)z & = & x(y z), \\
(fst)      & Fst[x,y]     & = & x, \\
(snd)      & Snd[x,y]     & = & y,  \\
(dpair)    & <x,y>z       & = & [xz,yz], \\
(ac)       & \varepsilon[\Lambda (x)y,z] & = & x[y,z], \\
(quote)    & (' x)y       & = & x,
\end{array}
\]
where $(dpair)$ connects the {\em pairing} and {\em coupling} operations,
$(ass)$ relates the {\em composition} and {\em application}.
An easy conclusion of $\$[x,y]z = \varepsilon [xz,yz]$ may be proved.
Hence, the manipulations with the combinators
$Fst$, $Snd$, and $\varepsilon$ become homogeneous.
Besides that, the equation
$$ (' M) = \Lambda(M \circ Snd) $$
is easy to verify giving rise to the equation $(' x)yz = xz$.
Now everything is prepared to set up an evaluation
in a cartesian closed category.

\begin{exm}[compiling the `categorical code'].
Let us apply both the semantic and syntactical rules
to evaluate the source-object
$F = (\lambda x.x[4,(\lambda x.x)3])+ = (\lambda.{\underline 0}$
                        $[4,(\lambda.{\underline 0})3])+ $:
\[
\begin{array}{lcl}
F' & = &                       \\
& = & \|F\|                      \\
& = & \|(\lambda.{\underline 0}[4,(\lambda.{\underline 0})3])+ \|       \\
& = & \$[\|\lambda.{\underline 0}[4,(\lambda.{\underline 0})3]\|,\|+\|]  \\
& = & \$[\Lambda(\|{\underline 0}[4,(\lambda.{\underline 0})3]\|),\|+\|]  \\
& = & \$[\Lambda(\$[0!,||[4,(\lambda.{\underline 0})3]\|)),\|+\|]  \\
& = & \$[\Lambda(\$[0!,<' 4,\$[\Lambda(0!), ' 3]>]),\Lambda(+ \circ Snd)].
\end{array}
\]
The final object gives the required categorical code.
\end{exm}

\begin{exm}[computing by closure].
To {\em compute by closure} means evaluate $F$ by applying $F'$
to the environment. Initially, for closed  $F$ an environment is empty,
thus $\rho = ()$. The strategy to evaluate $F'$ is to select the most
left and most inner expression. To save writing let to abbreviate:
$$ A = \$[0!,<' 4,B>],\ {\rm and}\ B = \$[\Lambda(0!),3]. $$
The chain of equations is as follows:
\[
\begin{array}{l}
[\Lambda(A),\Lambda(+ \circ Snd)]()  =   \\
 =  \varepsilon[\Lambda(A)(),\Lambda(+ \circ Snd)()] \\
 =  A \rho = \\
    {\rm (here:\ an\ abbreviation}\
  \rho = [(),\Lambda(+ \circ Snd)()]\ {\rm is\ applied)} \\
 =  \varepsilon[0!\rho,<' 4,B>\rho]  \\
 =  \varepsilon[\Lambda(+ \circ Snd)(),[' 4 \rho,B \rho]] \\
 =  \varepsilon[\Lambda(+ \circ Snd)(),[4,B \rho]] \\
 =  \varepsilon[\Lambda(+ \circ Snd)(),[4,\varepsilon[\Lambda(0!)\rho,
             ' 3 \rho]]] \\
 =  \varepsilon[\Lambda(+ \circ Snd)(),[4,\varepsilon[\Lambda(0!)\rho,3]]] \\
 =  \varepsilon[\Lambda(+ \circ Snd)(),[4,0![\rho,3]]] \\
 =  \varepsilon[\Lambda(+ \circ Snd)(),[4,3]]] \\
 =  (+ \circ Snd)[(),[4,3]] \\
 =  +(Snd[(),[4,3]] \\
 =  +[4,3]  \Rightarrow  7.
\end{array}
\]
\end{exm}

This result is the same as in case of direct computations given above.

\subsection{Avoiding Encapsulation: Supercombinators}

Now the process of compiling objects is under discussion.
The known approaches from the applicative computations give the
different (and distinct) strategies to transform one objects to other ones.
The discussion given above stimulates the useful intuition to mark
the specific features. The first approach deals with the direct
compiling the source-object into target-object using the
pre-specified set of combinators. Non optimized combinatory code
involves the set $\{{\cal I}, {\cal K}, {\cal S}\}$ as the basis
to compile-in. Note that the precise definitions are known before
the compiling has done. Another idea is to generate the combinators
during the compiling. In fact, the target set of resulting combinators
will be known after the compiling. The last strategy is
based on the specific objects called as {\em supercombinators}.
The supercombinator $\$S$ of arity $n$ is the $\lambda$-expression
($\lambda$-abstraction)
$\lambda x_1.\lambda x_2. \dots .\lambda x_n.E$ where $E$
is not abstraction. Thus, all the leading symbols of
abstraction `$\lambda$' bind exclusively $x_1, x_2, \dots, x_n$ and
the following restrictions are valid: \\
(1)~$\$S$ does not include any free variable; \\
(2)~every abstraction in $E$ is a supercombinator; \\
(3)~$n \ge 0$, so the symbols `$\lambda$' are not necessary.

It's a time to compare the different kinds of objects available
and their main computational properties. Replacing the free occurrences
of the formal parameters in the body of supercombinator
by the actual arguments is called as comprehension and is enforced
by $\beta$-conversion. Intuitively, the combinator is
the $\lambda$-abstraction that does not contain any free
occurrences of variables. Hence, some combinators are the supercombinators.
Similarly, some $\lambda$-expressions are the combinators.

\begin{exm} The objects $3$, $4$, $[3,4]+$, $\lambda x.x$
are the supercombinators. The objects
$\lambda x.y$ (free variable $y$),
$\lambda y. + y x$ (free variable $x$) are not supercombinators. \\
The object $\lambda f.f(\lambda x.f x 2)$ is a combinator
(all the variables are bound) but is not
a supercombinator (inner abstraction contains
the free variable $f$ violating the definition).
The combinators ${\cal I}, {\cal K}, {\cal S}$ are supercombinators.
Hence, the ${\cal IKS}$-compiling above
is based on supercombinators, in particular.
\end{exm}

The supercombinators of arity $0$ are called as
{\em constant applicative forms} (CAF), thus $(+\ 3\ 4)$ is a CAF.

\subsection{Compiling Objects with Supercombinators}

The actual programs contain the amount of abstractions.
Hence, the program is to be transformed to include the supercombinators
ultimately.
The notations for supercombinators are started
with the symbol `$\$$', e.g., $\$X = \lambda x.x$.
To stress the specific features of the supercombinators
rewrite this definition by $\$X x = x$.

The strategy selected is to transform
the compiled abstraction into: \\
({\em i}) the set of supercombinators' definitions; \\
({\em ii}) the evaluated expression.

This is depicted by the definitions of supercombinators
\begin{center}
\begin{minipage}[t]{60mm}
The definitions of supercombinators \\
. \dotfill .\\
. \dotfill .\\
\rule[1mm]{60mm}{.2mm}  \\
The evaluated expression
\end{minipage}
\end{center}

\begin{exm} The expression $([x].[y].-\ y\ x)3\ 4$
has the representation:
\begin{center}
\begin{minipage}[t]{40mm}
$\$ XY x\ y\ =\ -\ y\ x $ \\
\rule[1mm]{30mm}{.2mm}           \\
$\$XY 3\ 4$
\end{minipage}
\end{center}
\end{exm}

The compiling in supercombinators is straightforward.
\begin{exm}
Consider the example of a program:
$$ (\lambda x.x[4,(\lambda x.x)3]+.$$
(1)~Select the innermost abstraction,
     i.e. the abstraction that
     does not contain the other abstractions:
     $(\lambda x.x)$. It does not contain any free variable,
     thus it is the supercombinator named $\$X$.  \\
(2)~Include this object into the initial program: $(\lambda x.x[4,\$X 3])+$. \\
(3)~The innermost abstraction is the outermost abstraction,
    and it is the supercombinator $\$Y$ where $\$Y x = x[4,\$X 3]$. \\
(4)~The compiled code is:

\begin{center}
\begin{minipage}[t]{40mm}
$\$X x = x$            \\
$\$Y x = x[4,\$X 3] $   \\
\rule[1mm]{30mm}{.2mm}           \\
$\$Y +$
\end{minipage}
\end{center}

The resulting program is evaluated by reducing the supercombinators:
$$ \$Y + \Rightarrow +[4,\$X 3] \Rightarrow +[4,3] \Rightarrow 7. $$
\end{exm}

As we see, the particular solution may be generalized without difficulties.
Note that the example above was oversimplified.
More realistic programs include, at least, extraparameters and recursion.
This is known in programming with supercombinators,
and the $\lambda$-lifting procedure 
would assist
to optimized solutions.
In addition, the parameters are to be reordered
to save the amount of compiling steps. Note that
the multiple computations of the same subexpressions
may be avoided by the lazy-evaluation strategies.


\section{Final Remarks: Comparison of the Object Computations}

The main computational features are briefly outlined.

The compiling into supercombinators has some advantage:
the direct object view of computations. In addition,
the known side effects are avoided giving rise to the
purely functional computations.

The nature of $\lambda$-abstractions
attracts the researchers with their clear understanding
of the binding scopes for variables.

Combinatory representation brings a high degree
uniformity and homogeneity
into the domain of reasoning and computations
with the objects.

The particular systems of combinators, e.g.,
the categorical combinators are suitable
to separate the compiling process from
the computation itself leading to
the promising strategies of optimizing
the entire computation.

\subsection{Supersombinators}

This kind of objects gives an object-oriented solution
based on {\em combinator-as-object} doctrine with the possible
improvements of computations.

\begin{namelist}{{\tt (4)} {\it {   }}}
\item {(1)} Encapsulated objects are extracted from the body of
   the $\lambda$-abstractions.
\item {(2)} The optimization strategies are applicable to save
   the computational efforts.
\item {(3)} The object has a canonical representation, namely
    by supercombinator.
\item {(4)} Easy and natural ways to optimize the computations
    by the strategies.
\item {(5)} Easy to organize lazy evaluations.
\item {(6)} Mostly dynamic instructions-supercombinators.
     The instructions are generated during the compiling.
\item {(7)} The source-program is an arbitrary $\lambda$-expression.
\end{namelist}

\subsection{$\lambda$-Abstractions}

The $\lambda$-abstractions give a {\em function-as-object} notion.

\begin{namelist}{{\tt (4)} {\it {   }}}
\item {(1)} The objects contain both free and bound variables.
\item {(2)}  The binding ranges are easy observed.
\item {(3)}  The bound variables may be encoded by `nameless dummies'.
\item {(4)}  Not all the objects are encapsulated.
          Encapsulations are dominant.
\item {(5)}  The variety of initial objects.
\item {(6)}  Objects are reducible to combinators.
\item {(7)}  Direct computations by the rules of equational system.
\item {(8)}  The source-program is an arbitrary $\lambda$-expression.
\item {(9)}  Lazy evaluations need the additional strategies.
\end{namelist}

\subsection{Combinatory Representation}

This kind of an object representation generates the
{\em combinatory code} with the encapsulated objects.
Some objects are the objects-as-combinators.

\begin{namelist}{{\tt (4)} {\it {   }}}
\item {(1)} The objects contain only free variables as extraparameters,
     and bound variable are binding effects are avoided.
\item {(2)}  All the objects are encapsulated. Pure encapsulation.
\item {(3)}  The minimal initial supply of distinct objects.
\item {(4)}  The combinators are easy encoded by the $\lambda$-terms.
\item {(5)}  The source-program is an arbitrary $\lambda$-expression.
\item {(6)}  Some sets of combinators have the basis property.
\item {(7)}  Recursive programs are allowed.
\item {(8)}  Lazy evaluations need the additional strategies.
\end{namelist}

\subsection{Categorical Combinators}

The two aspects of computations are recently dominant.
First of them, and more traditional is based on
object-as-instruction with the possible combinatory
encoding. The second generates the computational models
relative to data models.

\begin{namelist}{{\tt (4)} {\it {   }}}
\item {(1)}  Fine theoretical framework.
\item {(2)}  Easy to optimize by the commutative diagrams.
\item {(3)}  Mostly the static combinators/instructions.
      The only dynamic instruction is application $\varepsilon$.
\item {(4)}  The initial supply of the objects/instructions is limited.
\item {(5)}  The source-program is an arbitrary $\lambda$-expression.
\item {(6)}  Recursive programs are allowed.
\item {(7)}  Lazy-evaluation strategies.
\item {(8)}  Encapsulation exists.
\end{namelist}




\section{Conclusions}

A large part of the difficulties with object notions
can be traced to there not being a conceptual framework.
A rough tracing of common object techniques
shares  distinct branches, - thus the
logical, categorical, and
mostly computational ones are outlined.

The essential ideas from the object calculi are based on a few
of initial concepts. The most important is a function which
is not determined by its domain and range but is determined
by the process. Those are combinators.
The combinators are the pure and elementary objects
which are combined by the metaoperator of application.

Another way to understand functions as objects (or: objects via functions)
is to use the functional abstraction. This is a metaoperator
which is added to application.

The computational systems based on applications and abstractions
are referred as applicative computational systems (ACS).
ACS to the contrast with the operator,
or imperative computational systems, have some important advantages.
Among them is the clear mathematical foundation.
The usefulness of the mathematical properties is even more than
a theory of computations.

The preliminary study to apply ACS to the domain of objects
as they are in a database theory shows the following:

\begin{namelist}{{\tt (4)} {\it {   }}}
\item {(1)} Modeling the objects and corresponding
        computations for combinatory logic and $\lambda$-calculus
        involves a domain of objects and a set of (meta-)operations
        that are to be represented by the elements of the domain.
\item {(2)} The class of possible operations depends on the
        particular constructions: the {\em definable}
        operations should be represented.
\item {(3)} To some extent the study of different object notions
        can be pursued independently of the particular
        data models.
\end{namelist}

\section{Acknowledgements}

The author is indebted to Prof. Alexandre Kuzichev (MSU) for
drawing my attention to the computational properties of the
objects.


\nocite{Brod:95}
 \nocite{NeRo:95}
\nocite{EhGoSe:93}


\addcontentsline{toc}{section}{References}
\newcommand{\noopsort}[1]{} \newcommand{\printfirst}[2]{#1}
  \newcommand{\singleletter}[1]{#1} \newcommand{\switchargs}[2]{#2#1}


\end{document}